\documentclass[prd,nofootinbib,superscriptaddress]{revtex4-2} 
\pdfoutput=1 

\usepackage{amsmath, amssymb, amsthm, graphicx, epsfig, fancyhdr,epsfig, slashed, mathrsfs}

\usepackage{tikzsymbols}
\usepackage{natbib}
\usepackage{float}
\usepackage{booktabs}
\usepackage{bm}         
\usepackage{array}
\usepackage{mathtools} 
\usepackage{tikz,xcolor,hyperref}

\usepackage{txfonts}

\definecolor{lime}{HTML}{A6CE39}
\DeclareRobustCommand{\orcidicon}{
	\begin{tikzpicture}
	\draw[lime, fill=lime] (0,0) 
	circle [radius=0.2] 
	node[white] {{\fontfamily{qag}\selectfont \tiny ID}};
	\draw[white, fill=white] (-0.0625,0.095) 
	circle [radius=0.007];
	\end{tikzpicture}
	\hspace{-2mm}
}

\foreach \x in {A, ..., Z}{\expandafter\xdef\csname orcid\x\endcsname{\noexpand\href{https://orcid.org/\csname orcidauthor\x\endcsname}
			{\noexpand\orcidicon}}
}

\newcommand{\be}{\begin{equation}}
\newcommand{\ee}{\end{equation}}
\newcommand{\bea}{\begin{eqnarray}}
\newcommand{\eea}{\end{eqnarray}}

\begin{document}

\title{Yukawa theory in non-perturbative regimes: \\ 
\it{towards confinement, exact $\beta$-function and conformal phase}}

\author{Marco Frasca\orcidA{}}
\email{marcofrasca@mclink.it}
\affiliation{Rome, Italy}

\author{Anish Ghoshal\orcidB{}}
\email{anish.ghoshal@fuw.edu.pl}
\affiliation{Institute of Theoretical Physics, Faculty of Physics, University of Warsaw, ul. Pasteura 5, 02-093 Warsaw, Poland}

\begin{abstract}
We study possible hints towards confinement in a Z$_2$-invariant Yukawa system with massless fermions and a real scalar field in the strongly-coupled regime. Using the tools developed for studying non-perturbative physics via Jacobi elliptical functions, for a given 
 but not unique choice of the vacuum state, we find the exact Green's function for the scalar sector so that, after integrating out the scalar degrees of freedom, we are able to recover the low-energy limit of the theory that is a fully non-local Nambu-Jona-Lasinio (NJL) model. We provide an analytical result for the Renormalization Group (RG) running of the self-interaction coupling in the scalar sector and critical indexes in the strongly-coupled regime. In the fermion sector, we provide some clues towards confinement, after deriving the gap equation with the non-local NJL model, a property which is well-known to not emerge in the local limit of this model. We conclude that, for the  scalar-Yukawa theory in the non-perturbative domain with our choice of the vacuum state, the fundamental fermions of the theory form bound states and cannot be observed as asymptotic states.
\end{abstract}

\maketitle

\section{Introduction}
\label{Sect:intro}

The Standard Model (SM) of particle physics is a successful theory that has passed a number of phenomenological high-precision tests. One crucial building block is the 
%
scalar sector which renders the perturbative expansion of correlation functions well-defined and parameterizes the masses of matter fields and weak gauge bosons in phenomenologically consistent way. So far, the Higgs sector has been tested by precision measurements of single Higgs boson production and decay channels (for current status of Higgs phenomenology, see e.g.~Refs.~\cite{Workman:2022ynf,ATLAS:2022vkf} and references therein), and will eventually be probed at the LHC and future colliders through a measurement of Higgs self-interaction in di-Higgs production (for recent LHC constraints, see Ref.~\cite{ATLAS:2022jtk}). 

Beyond the unmatched phenomenological success of the SM at the reach of the current experiments, the triviality and hierarchy problems are among the intricate problems of the SM framework that have been particularly inspiring the research community to design new ideas for a larger fundamental framework. Whereas the hierarchy problem is not a fundamental problem in the sense of rendering the SM ill-defined, the triviality problem truly inhibits an extension of the SM to arbitrarily high momentum scales. Loosely speaking, the scale of maximum ultra-violet (UV) extension $\Lambda_{\text{UV,max}}$ induced by triviality is related to the critical indices of the theory that characterize its universality class. For instance, the 3+1D $\phi^4$-theory seems to belong to the same universality class of the 4D Ising model. In a recent paper, Aizenman and Duminil-Copin \cite{Aizenman:2019yuo} have proven marginal triviality of the 
latter.
This proof relies on a trivial ground state for the theory where the 1P-correlation function is taken to be a constant. We say ``marginal triviality'' as some differences are seen in the scalar field theory with respect to the Ising model as we will discuss below. In the SM, the Higgs potential of
$\lambda\phi^4$/4
type is renormalizable and produces finite scattering amplitudes and cross sections. However, renormalization demands that the cutoff scale $\Lambda_C$ must be taken to infinity, $\Lambda_C\rightarrow\infty$~\cite{Dyson1949,Dyson1952}. Then, the renormalized Higgs quartic coupling vanishes, $\lambda \to 0$, which is known as the triviality problem of the SM~\cite{Landau1955,LandauPomeranchuk1955,Landau1956,Callaway1986,Callaway1988,Wilson}. Note that, the triviality issue could be addressed if the Standard Model can be asymptotically safe \cite{Mann:2017wzh,Antipin:2018zdg,Molinaro:2018kjz,Wang:2018yer,Sannino:2019sch,Cacciapaglia:2023ghp,Cacciapaglia:2020qky,Christiansen:2017gtg}.

Such a triviality problem exists in both the Higgs sector \cite{Wilson:1973jj,Luscher:1987ek,Hasenfratz:1987eh,Heller:1992js,Callaway:1988ya,Rosten:2008ts} as well as in the U(1) gauge sector \cite{Gell-Mann:1954yli,Gockeler:1997dn,Gies:2004hy}. But since $\Lambda_{\text{UV, max}}$ of the Higgs sector is much smaller than that of the U(1) sector \cite{Gies:2004hy}, evading triviality in the Higgs sector is of primary importance. In fact, for large Higgs boson masses, the scale $\Lambda_{\text{UV,max}}$ could be much smaller than the Planck or Grand unification scales \cite{Cabibbo:1979ay,Kuti:1987nr,Hambye:1996wb,Fodor:2007fn,Gerhold:2008mb}. For supersymmetric models, $\Lambda_{\text{UV,max}}$ can even be smaller than in the SM \cite{Yndurain:1991vm}. 

In this work, we demonstrate how triviality appears in the scalar sector in a rather peculiar way: The 
K\"all\'en--Lehmann
representation of the propagator does not display any bound state. We have just a harmonic oscillator spectrum. This could change due to interactions with a fermion field. In such a case, non-trivial physics arises in the fermion sector, with a possible confinement phenomenon and the gap equation, after the scalar degrees of freedom are integrated out. Therefore, a trivial quantum field that interacts with other matter fields could yield some non-trivial observable physics as already seen for the SM at the LHC.

A first-principle derivation of confinement and its space-time dynamics causing the absence of fundamental degrees of freedom in asymptotic states of the theory is yet another long-standing theoretical challenge in particle physics (for a recent review of this problem, see e.g.~Ref.~\cite{Pasechnik:2021ncb} and references therein). Difficulties arise as we are generally unable to treat QCD in the strong coupling regime in the framework of the standard local quantum field theory. Among the modern approaches to confinement are the lattice computations and effective field theories that are generally not derivable from the microscopic field theory of QCD. Within the scope of the latter, the low-energy limit of QCD has been studied with local and non-local Nambu--Jona-Lasinio (NJL) models~\cite{Nambu:1961tp,Nambu:1961fr,Klevansky:1992qe,GomezDumm:2006vz,Hell:2008cc}. The local version of the NJL-model, as initially conceived in Refs.~\cite{Nambu:1961tp,Nambu:1961fr}, does not exhibit confining properties. For bound states obtained after bosonization~\cite{Ebert:1997fc}, there is a threshold for their decay into quark and anti-quark pairs, appearing as free asymptotic states of the theory, that have never been observed. Non-locality can help to resolve such an issue in a consistent manner~\cite{Bowler:1994ir}. In this paper, we will demonstrate how a generalized, non-local NJL model emerges in the strong coupling regime of the considered Yukawa theory, along with its possible confining properties according to the criteria given in Refs.~\cite{Bowler:1994ir,Roberts:1994dr}. The idea is to consider a 
Beyond Standard Model (BSM)
scalar sector interacting with a Dirac fermion in a non-perturbative regime where confinement can emerge.

In order to study the non-perturbative regime in the Yukawa theory, we utilize the exact solutions found in terms of the Jacobi elliptical functions following the analytic approach of Dyson-Schwinger equations originally devised by Bender, Milton and Savage in Ref.~\cite{Bender:1999ek}. In this case, the Green's functions of the theory are represented analytically, and therefore it is straightforward to understand the effect of the background on the interactions that remain valid even in the strongly-coupled regime \cite{Frasca:2015yva}. This technique has been recently applied to QCD in Refs.~\cite{Frasca:2021yuu,Frasca:2021mhi,Frasca:2022lwp,Frasca:2022pjf,Chaichian:2018cyv} and to the SM Higgs sector in Ref.~\cite{Frasca:2015wva}, as well as to other types of models over the past two decades in Refs.~\cite{Frasca:2019ysi, Chaichian:2018cyv, Frasca:2017slg, Frasca:2016sky, Frasca:2015yva, Frasca:2015wva, Frasca:2013tma, Frasca:2012ne, Frasca:2009bc, Frasca:2010ce, Frasca:2008tg, Frasca:2009yp, Frasca:2008zp, Frasca:2007uz, Frasca:2006yx, Frasca:2005sx, Frasca:2005mv, Frasca:2005fs}. Recently, some of the authors have employed this technique to study non-perturbative hadronic contributions to the muon anomalous magnetic moment (g-2)$_{\mu}$ \cite{Frasca:2021yuu}, QCD in the non-perturbative regime \cite{Frasca:2021mhi,Frasca:2022lwp,Frasca:2022pjf}, non-perturbative false vacuum decay \cite{Frasca:2022kfy}, as well as to explore the mass gap and confinement in string-inspired infinite-derivative and Lee-Wick theories \cite{Frasca:2020jbe,Frasca:2020ojd,Frasca:2021iip}.

It is important to emphasize that our study pertains to a given choice of the vacuum state that boils down to a given solution for the one-point correlation function of the scalar sector. This by no means entails that other choice either are not acceptable or are superseded by ours. 
Instead,
we consider interesting to see how non-trivial non-perturbative physics, like confinement and a mass gap, can emerge by this particular choice that grants analytical solutions in closed form.

Given the above {\it proviso}, in this paper we investigate non-perturbative properties of a simple Yukawa theory and show a derivation of the mass gap for a massless scalar sector, a gap equation for the fermion and the exact beta function in the strongly-coupled regime for the scalar field with a correction due to a small Yukawa coupling to the fermion. We will demonstrate that these are fully consistent with expectations from triviality for the scalar sector and  provide important hints for confinement for the fermion sector. 

The paper is organized as follows. In Sec.~\ref{Sect:DS}, we discuss the model and the technique we use to solve it. We show the first correlation functions of the scalar sector and the Dyson-Schwinger equations for the scalar sector interacting with a fermion. The Renormalization Group (RG) behavior of the scalar sector is discussed through the technique of exact functional renormalization group. In Sec.~\ref{Sect:FF}, we get the next-to-leading order correction to the spectrum of the scalar theory arising form the Yukawa-coupled fermion in the weak coupling limit. In Sec.~\ref{Sect:PF}, we give explicitly the partition function in a next-to-leading order approximation and integrate out the scalar contribution obtaining the non-local NJL model. We bosonize this model to show the true spectrum of the theory made by bound states of the fermion field and derive the gap equation for the latter. In Sec.~\ref{Sect:GE}, we solve the gap equation both in the local and non-local case displaying some hints of confinement. In Sec.~\ref{Sect:CC}, we present our conclusions.

\section{Dyson-Schwinger equations of Yukawa theory}
\label{Sect:DS}

\subsection{Scalar sector}

Let us start by introducing the basics of the analytic non-perturbative approach of Dyson-Schwinger equations~\cite{Bender:1999ek,Frasca:2015yva}. For this purpose, we start by considering the following Lagrangian \cite{Frasca:2015yva}
for a quartic scalar field $\phi$
\be
{\cal L} = \frac{1}{2}(\partial\phi)^2-\frac{\lambda}{4}\phi^4
\ee
and show how $\phi$ acquires a dynamical mass by means of its quartic interaction. 
Here $\lambda$ is the coupling of the field.
We wish to solve the quantum equation of motion,
\begin{equation}
    \partial^2\phi+\lambda\phi^3=j \,,
\end{equation}
where $j$ is an arbitrary current,
given the generating functional in the form
\begin{equation}
    Z[j]=\int[d\phi]e^{i\int d^4x\left[\frac{1}{2}(\partial\phi)^2-\frac{\lambda}{4}\phi^4+j\phi\right]} \,.
\end{equation}
From this expression, the correlation functions can be computed by means of Dyson-Schwinger equations (DSEs). Namely, the DSE for the one-point function reads (a recent derivation of this set of equations can be found in e.g. \cite{Frasca:2022kfy})
\begin{equation}
\label{eq:g10}
   \partial^2 G_1(x)+\lambda\left([G_1(x)]^3+3G_2(x,x)G_1(x)+G_3(x,x,x)\right)=0 \,,
\end{equation}
while for the two-point function 
\begin{equation}
\label{eq:g20}
   \partial^2G_2(x,y)+\lambda\left(3[G_1(x)]^2G_2(x,y)+3G_2(x,x)G_2(x,y)
	+3G_3(x,x,y)G_1(x)+G_4(x,x,x,y)\right)=\delta^4(x-y) \,.
\end{equation}
One can explicitly demonstrate that, while the background solution $G_1(x)$ is a non-trivial function of the spacetime coordinate $x$ and hence formally violates translation invariance, all the observables naturally appear to be translationally-invariant. We will discuss this point below. 

Our chosen solution for Eq.~(\ref{eq:g10}), that is our representation of the vacuum of the theory, can be written as the so-called Fubini instanton \cite{Fubini:1976jm,Lipatov:1976ny}
\begin{equation}
\label{solG1}
    G_1(x)=\sqrt{\frac{2\mu^4}{m_\phi^2+\sqrt{m_\phi^4+2\lambda\mu^4}}}{\rm sn}\left(p\cdot x+\chi\left|\frac{-m_\phi^2+\sqrt{m_\phi^4+2\lambda\mu^4}}{-m_\phi^2-\sqrt{m_\phi^4+2\lambda\mu^4}}\right.\right) \,,
\end{equation}
where $\mu$ and $\chi$ are arbitrary integration constants (dimension-one energy scale and dimensionless phase, respectively), 
sn$(\xi,k)$
is a Jacobi elliptic function and, provided we define $m_\phi^2 \equiv 3\lambda G_2(0)$, $G_3(x,x,x)=0$ (see the appendix to see the consistency of this choice), the four-momentum $p$ of the quasi-particle satisfies 
\begin{equation}
\label{eq:disp}
    p^2=m_\phi^2+\frac{\lambda\mu^4}{m_\phi^2+\sqrt{m_\phi^4+2\lambda\mu^4}} \,.
\end{equation}
It must be emphasized that the momentum $p$ entering into the dispersion relation arises by the integration of the equation for $G_1$ and does not correspond to the momentum of asymptotic states in quantum field theory. Once again we point out that this choice is neither unique nor we are able to prove that it is the best one. We choose it because is in a closed analytical form, all other results we will obtain from it are too and some interesting understanding of non-perturbative physics can be straightforwardly obtained. 

It is important to notice that $G_2(0)$ gives rise to the dynamical mass $m_\phi^2$ while the dispersion relation (\ref{eq:disp}) would be massive even for $G_2(0)=0$. The solution $G_1(x)$ depends explicitly on space-time points and thus it formally violates translation invariance of the action. The latter is a generic feature of the dynamical ground-state, also found in other approaches beyond the Fubini instanton \cite{Fubini:1976jm,Lipatov:1976ny} (for a comprehensive discussion of non-trivial properties of the ground state e.g.~in quantum non-abelian gauge theories, see Refs.~\cite{Batalin:1976uv,Savvidy:1977as,Pagels:1978dd,Olesen:1980nz,Shuryak:1983ni,Pasechnik:2013sga,Pasechnik:2013poa,Addazi:2018fyo,Addazi:2018ctp}, as well as recent reviews of Refs.~\cite{Pasechnik:2016sbh,Addazi:2022whi,Pasechnik:2021ncb} and references therein). However, the vacuum correlation function $G_1(x)$ is not an observable of the theory by itself. It can be shown explicitly that despite of a non-trivial $x$-dependence of $G_1(x)$, the translation invariance is recovered
for
the propagator $G_2$ as we will see below. This grants a proper symmetric behavior to the observables of the theory based on LSZ theorem. An important example of this fundamental feature of a quantum theory can be found in Ref.~\cite{Addazi:2018fyo}, where it has been demonstrated that quasi-periodic singularities in the homogeneous time-dependent Yang-Mills condensate result in a translationally-invariant ground-state energy density in the form of cosmological constant.

It is then straightforward to obtain the two-point correlation function in momentum space \cite{Frasca:2013tma}. If the mass shift due to renormalization could be neglected (below, we will consider the full formula to a proper evaluation of the RG behavior of the scalar sector),
the propagator is given by
\begin{equation}
\label{eq:G2}
   G_2(p)=\Delta(p)=\frac{\pi^3}{4K^3(-1)}
	\sum_{n=0}^\infty\frac{e^{-(n+\frac{1}{2})\pi}}{1+e^{-(2n+1)\pi}}(2n+1)^2\frac{1}{p^2-m_n^2+i\epsilon} \,,
\end{equation}
and one obtains the mass spectrum
\begin{equation}
\label{eq:ms}
   m_n=(2n+1)\frac{\pi}{2K(-1)}\left(\frac{\lambda}{2}\right)^\frac{1}{4}\mu \,,
\end{equation}
where $K(-1)=1.31102877714\ldots$ is the numerical value of the complete elliptical integral of the first kind\footnote{We use the notation of $K(-1)$ for the complete elliptic integral of the first kind since we work with the square of the modulus $\kappa$, that is we consider $K(\kappa^2)$ as for all the elliptic functions and relative integrals that we consider in this paper. This choice has been adopted by some computer algebra systems like Wolfram Mathematica$^\copyright$. Another notation is also possible using the modulus itself and, in such a case, one should write $K(i)$ having identical numerical value as given in the main text.}. This representation for $G_2$ will solve the equation (\ref{eq:g20}) and is manifestly translation invariant, provided we truncate the infinite system of DSEs by assuming $G_3(x,x,y)=0$ and $G_4(x,x,x,y)=0$ as we show in the appendix.

Some considerations are needed for the $Z_2$ symmetry. We started with a Lagrangian that is $Z_2$-invariant that is, physics does not change by the transformation $\phi\rightarrow -\phi$. Our solution breaks this symmetry and, indeed, the theory has a zero mode. This is well explained in Ref.\cite{Frasca:2022kfy}. Therefore, we do not expect the vanishing of any higher-order nP-correlation function on the basis of the $Z_2$ symmetry.
%

We emphasize that this is an exact solution to the set of Dyson-Schwinger 
%
equations. Indeed, through the choice to set all higher-order correlation functions
%
to zero when evaluated to equal points is equivalent to get all the correlation functions expressed in closed analytical form through $G_1$ and $G_2$. This is a Gaussian exact solution where the cumulant expansion in the path integral is that of a Gaussian distribution in Euclidean metric. We just prove that it exists.
%

\subsection{Adding fermions}

Let us now extend the scalar theory above by adding fermions such that
\be   
L_y=\frac{1}{2}(\partial\phi)^2-\frac{\lambda}{4}\phi^4+{\bar q}({\slashed\partial}-Y_f\phi-m_q)q \,,
\ee
where $\phi$ is the scalar field, $q$ is the fermion field with mass $m_q$, and $Y_f$ is the corresponding Yukawa coupling. Therefore, our partition function will have the form
\be
Z[j,\bar\eta,\eta]=\int[d\phi][d{\bar q}][dq]e^{i\int d^4x\left(L_y+j\phi+\bar\eta q+{\bar q}\eta\right)} \,,
\ee
where $j,\ \eta,\ \bar\eta$ are the currents. We can introduce the following correlation functions, some of them come out from mixed derivatives with respect to $j$, $\eta$ and $\bar\eta$,
\bea
\label{eq:def}
q_1(x)&=&-i\left.\frac{1}{Z}\frac{\delta Z}{\delta\bar\eta(x)}\right|_{j,\bar\eta,\eta=0}, \nonumber \\
S_q(x,x')&=&-i\left.\frac{\delta {q}_1(x)}{\delta\bar\eta(x')}\right|_{j,\bar\eta,\eta=0} \,,
\nonumber \\
Q(x,x')&=&-i\left.\frac{\delta {q}_1(x)}{\delta j(x')}\right|_{j,\bar\eta,\eta=0}, \nonumber \\
W_q(x,x')&=&-i\left.\frac{\delta {G}_1(x)}{\delta\bar\eta(x')}\right|_{j,\bar\eta,\eta=0} \,.
\eea
%
Then, the Bender-Milton-Savage method yields the following system of DSEs for one-point functions, the scalar $G_1(x)$ and fermion $q_1(x)$,
\begin{eqnarray}
      &&\partial^2G_{1}(x)+\lambda(G_{3}(x,x,x)
		+3G_{2}(x,x)G_{1}(x)+G_1^3(x)) \nonumber \\
		&&=Y_f\sum_{q}S_{q}(x,x)+Y_f\sum_{q}{\bar q}_1(x)q_1(x) \,, \nonumber \\
	&& (i\slashed\partial-m_q)q_1(x)+Y_fG_1(x) q_1(x)+Y_fW_q(x,x)= 0 \,.  
\end{eqnarray}
Here, $S_q$ is the fermion propagator defined below and $q_1$ is the 1P-correlation function for the fermion (note that ${\bar q}_1(x)=q_1^*(x)\gamma^0$). Similar DSEs can be obtained, for instance, from such a system in a gauge theory such as QCD, previously discussed in Ref.~\cite{Frasca:2021mhi}, switching consistently from a gauge field to a scalar field. As usual for the infinite set of coupled 
Dyson-Schwinger equations,
the equations for the lower-order correlation functions depend on the higher-order ones, and practically useful results can be obtained by a consistent truncation of this system at a certain order by setting higher-order functions to zero only when two or more arguments in such correlation functions coincide.

Taking the next step, the corresponding equations for two-point functions are
\begin{eqnarray}
\label{eq:DS1}
&&\partial^2G_2(x,y) + 3\lambda G_2(x,x)G_2(x,y) + 
3\lambda G_1^2(x)G_2(x,y) \nonumber \\
&&=Y_f\sum_{q}{\bar Q}(x,y)q_1(x) + Y_f\sum_{q}{\bar q}_1(x)Q(x,y) + 
\delta^{(4)}(x-y) \,, \nonumber \\
&&(i\slashed\partial-m_q)S_q(x,y) + Y_fG_1(x) S_q(x,y) = \delta^{(4)}(x-y) \,, \nonumber \\  
&&\partial^2W_q(x,y) + 3\lambda G_2(x,x)W_q(x,y) + 3\lambda G_1^2(x)W_q(x,y) = 
Y_f{\bar q}_1(x)S_q(x,y) \,, \nonumber \\
&&(i\slashed\partial-m_q)Q(x,y)+Y_fG_1(x)Q(x,y)+Y_f\Delta_\phi(x,y)q_1(x)=0 \,.
\end{eqnarray}
where $G_2(x,y)$ the 2P-correlation function of the scalar field discussed in the preceding section but taking into account fermionic contributions and $G_1(x)$ the 1P-correlation function of the scalar field. Similarly, $Q(x,y)$ and $W_q(x,y)$ are defined in (\ref{eq:def}). We recognize from this set of equations that a pure scalar-field solution works if we take the ordering $\lambda\gg Y_f$. 
%

\subsection{Functional renormalization group analysis of the scalar sector}
%

In order to have a proper understanding of the behavior of the scalar field with respect the running coupling, we exploit the technique of the exact functional renormalization group. We show that this agrees fairly well with our previous discussion.

We can write the Wetterich equation for the Effective Average Action (EAA) as \cite{Wetterich:1992yh} 
\be
k\partial_k\Gamma_k=\frac{1}{2}\operatorname{Tr}\left[\left(\frac{\delta^2\Gamma_k}{\delta\phi\delta\phi}+R_k\right)^{-1}k\partial_kR_k\right].
\ee
In our case, 
using Euclidean metric, the effective action is given by
\cite{Frasca:2022kfy,Calcagni:2022tls,Calcagni:2022gac} (we have added a subscript $k$ on the running constants of the theory)
\be
\Gamma_k[\phi]=\int d^4x\left(\frac{1}{2}(\partial\phi)^2+\frac{\lambda_k}{4}\phi^4
+\frac{1}{2}\sqrt{2\lambda_k}\mu_k^2(\phi-\phi_{0k})^2-\frac{1}{2}(2\lambda_k)^\frac{3}{4}\mu_k\operatorname{sn}(\theta_k|-1)(\phi-\phi_{0k})^3+\frac{1}{2}\phi R_k\phi
\right).
\ee
%
%
As pointed out in Ref.~\cite{Frasca:2013tma}, the real dependence of higher order correlation functions on $\lambda$ is decreasing at increasing coupling. In a physical situation where such a coupling is seen to run to higher values at increasing energies, we expect higher order corrections arising from vertexes to decrease with it and so, further corrections to the effective potential for false vacuum should be neglected. We will discuss the running coupling below.
Then,
($\partial_t=k\partial_k$)
\bea
&&\partial_t\Gamma_k[\phi]=
\int d^4x\left[\frac{\partial_t\lambda_k}{4}\phi^4
+\frac{1}{4}\sqrt{\frac{2}{\lambda_k}}\partial_t\lambda_k\mu_k^2(\phi-\phi_{0k})^2
+\sqrt{2\lambda_k}\mu_k\partial_t\mu_k(\phi-\phi_{0k})^2\right. \\
&&-\sqrt{2\lambda_k}\mu_k^2(\phi-\phi_{0k})\partial_t\phi_{0k}
-\frac{3}{8}2^\frac{3}{4}\lambda_k^{-\frac{1}{4}}\partial_t\lambda_k\mu_k\operatorname{sn}(\theta_k|-1)(\phi-\phi_{0k})^3
-\frac{1}{2}(2\lambda_k)^\frac{3}{4}\partial_t\mu_k\operatorname{sn}(\theta_k|-1)(\phi-\phi_{0k})^3 \nonumber \\
&&\left.-\frac{1}{2}(2\lambda_k)^\frac{3}{4}\mu_k\operatorname{cn}(\theta_k|-1)\operatorname{dn}(\theta_k,i)\partial_t\theta_k(\phi-\phi_{0k})^3
+\frac{3}{2}(2\lambda_k)^\frac{3}{4}\mu_k\operatorname{sn}(\theta_k|-1)(\phi-\phi_{0k})^2\partial_t\phi_{0k}
\right].
\eea
Similarly,
\be
\frac{\delta\Gamma_k}{\delta\phi\delta\phi}=-\partial^2+3\lambda_k\phi^2+\sqrt{2\lambda_k}\mu_k^2-3(2\lambda_k)^\frac{3}{4}\operatorname{sn}(\theta_k|-1)\mu_k(\phi-\phi_{0k}).
\ee
This yields
\bea
&&\int d^4x\left[\frac{\partial_t\lambda_k}{4}\phi^4
-\sqrt{2\lambda_k}\mu_k^2\partial_t\phi_{0k}(\phi-\phi_{0k})
+\left(\frac{1}{4}\sqrt{\frac{2}{\lambda_k}}\partial_t\lambda_k\mu_k^2
+\sqrt{2\lambda_k}\mu_k\partial_t\mu_k
+\frac{3}{2}(2\lambda_k)^\frac{3}{4}\mu_k\operatorname{sn}(\theta_k|-1)\partial_t\phi_{0k}\right)(\phi-\phi_{0k})^2
\right. \nonumber \\
&&
\left.
-\left(\frac{3}{8}2^\frac{3}{4}\lambda_k^{-\frac{1}{4}}\partial_t\lambda_k\mu_k\operatorname{sn}(\theta_k|-1)
+\frac{1}{2}(2\lambda_k)^\frac{3}{4}\partial_t\mu_k\operatorname{sn}(\theta_k|-1)+\frac{1}{2}(2\lambda_k)^\frac{3}{4}\mu_k\operatorname{cn}(\theta_k|-1)\operatorname{dn}(\theta_k|-1)\partial_t\theta_k
\right)
(\phi-\phi_{0k})^3
\right] \nonumber \\
&&=\operatorname{Tr}\left[\frac{k\partial_kR_k}{-\partial^2+R_k+3\lambda_k\phi^2+\sqrt{2\lambda_k}\mu_k^2-3(2\lambda_k)^\frac{3}{4}\operatorname{sn}(\theta_k|-1)\mu_k(\phi-\phi_{0k})}\right], \nonumber \\
&&=\operatorname{Tr}\left[k\partial_kR_kG_{0k}\frac{1}{1-3(2\lambda_k)^\frac{3}{4}\operatorname{sn}(\theta_k|-1)\mu_kG_{0k}(\phi-\phi_{0k})}\right].
\eea
We have to expand both sides around $\phi=\phi_{0k}$.
Thus,
\bea
&&\int d^4x\left[-\frac{\partial_t\lambda_k}{4}\phi_{0k}^4
+\left(\sqrt{2\lambda_k}\mu_k^2\partial_t\phi_{0k}-\partial_t\lambda_k\phi_{0k}^3\right)(\phi-\phi_{0k})\right. \nonumber \\
&&-\left(\frac{1}{4}\sqrt{\frac{2}{\lambda_k}}\partial_t\lambda_k\mu_k^2
+\sqrt{2\lambda_k}\mu_k\partial_t\mu_k
+\frac{3}{2}(2\lambda_k)^\frac{3}{4}\mu_k\operatorname{sn}(\theta_k|-1)\partial_t\phi_{0k}+\frac{3\partial_t\lambda_k}{2}\phi_{0k}^2\right)(\phi-\phi_{0k})^2
\nonumber \\
&&+\left(\frac{3}{8}2^\frac{3}{4}\lambda_k^{-\frac{1}{4}}\partial_t\lambda_k\mu_k\operatorname{sn}(\theta_k|-1)
+\frac{1}{2}(2\lambda_k)^\frac{3}{4}\partial_t\mu_k\operatorname{sn}(\theta_k|-1)+\frac{1}{2}(2\lambda_k)^\frac{3}{4}\mu_k\operatorname{cn}(\theta_k|-1)\operatorname{dn}(\theta_k|-1)\partial_t\theta_k
-\partial_t\lambda_k\phi_{0k}\right)
(\phi-\phi_{0k})^3 \nonumber \\
&&\left.
-\frac{\partial_t\lambda_k}{4}(\phi-\phi_{0k})^4
\right] \nonumber \\
&&=\operatorname{Tr}\left[k\partial_kR_k{\tilde G}_{0k}\right]+O(\phi-\phi_{0k}),
\eea
where
\be
{\tilde G}_{0k}=\frac{1}{-\partial^2+R_k+\sqrt{2\lambda_k}\mu_k^2+3\lambda_k\phi_{0k}^2}.
\ee
Equating the terms of the expansion for powers of $\phi-\phi_{0k}$ yields the first 
%
RG equation for the running of the coupling
\be
\label{eq:rc}
\partial_t\lambda_k=4\operatorname{Tr}(k\partial_kR_k{\tilde G}_{0k})\left(\int d^4x\phi_{0k}^4\right)^{-1}.
\ee
We have
\be
\phi_{0k}(x)=\mu_k(2/\lambda_k)^\frac{1}{4}\operatorname{sn}(p\cdot x+\theta_k|-1).
\ee
To evaluate the volume integral in Eq.(\ref{eq:rc}), we select a finite volume $V$ and exploit the arbitrariness in the parameters of Eq.(\ref{solG1}) here properly relabeled, with the choice ${\bm p}=0$, so that
\be
\phi_{0k}(x)=\mu_k(2/\lambda_k)^\frac{1}{4}\operatorname{sn}\left(\mu_k(\lambda_k/2)^\frac{1}{4}t+\theta_k|-1\right),
\ee
and we have the formula for a very large but finite volume
\be
\int d^4x\phi_{0k}^4=\frac{4}{3}V\mu_k^4\lambda_k^{-1}.
\ee
If we choose the optimal regulator $R_k(p)=(k^2-p^2)\theta(k^2-p^2)$ and $\partial_kR_k(p)=2k\theta(k^2-p^2)$ and
\be
{\tilde G}_{0k}(p)=\sum_{n=0}^\infty\frac{B_n}{p^2+R_k(p)+(2n+1)^2m_{0k}^2+\delta m^2_k},
\ee
where $m_{0k}=(\pi/2K(i))(\lambda_k/2)^\frac{1}{4}\mu_k$ and we have set $\delta m^2_k=\sqrt{2\lambda_k}\mu_k^2$, we get
\be
\partial_t\lambda_k=\frac{3\lambda_k}{\mu_k^4}\int\frac{d^4p}{(2\pi)^4}\sum_{n=0}^\infty B_n\frac{2k^2\theta(k^2-p^2)}{p^2+(k^2-p^2)\theta(k^2-p^2)+(2n+1)^2m_{0k}^2+\delta m^2_k}.
\ee
The integral can be easily performed 
%
giving
\be
\partial_t\lambda_k=\frac{3\lambda_k}{\mu_k^4}\frac{\Omega_4}{2(2\pi)^4}k^6\sum_{n=0}^\infty\frac{B_n}{k^2+(2n+1)^2m_{0k}^2+\delta m^2_k}=
\frac{3\lambda_k}{16\pi^2}\frac{k^6}{\mu_k^4}\sum_{n=0}^\infty\frac{B_n}{k^2+(2n+1)^2m_{0k}^2+\delta m^2_k}.
\ee
%
We recognize that, for $\mu_k=k/\lambda_k^\frac{1}{4}$ (it should be remembered that this is just an arbitrary integration constant)\footnote{It should be noticed that, with this choice of the integration constant $\mu$, the dispersion relation takes the form $p^2=\mu^2/\sqrt{2}$. This choice, driven by the Functional Renormalization Group (FRG), will not be implemented by us in the rest of the paper.}.
%
We finally get the $\beta$-function
\be
\partial_t\lambda_k=\kappa_0\frac{3\lambda_k^2}{16\pi^2},
\ee
where
\be
\kappa_0=\sum_{n=0}^\infty\frac{B_n}{1+(2n+1)^2(\pi/2K(i))^2(1/2)^\frac{1}{2}+2^\frac{1}{2}}\approx 1.62802.
\ee
This is essentially the Wilson-Fisher $\beta$-function \cite{Peskin:1995ev}.

\section{Corrections from Yukawa interaction}
\label{Sect:FF}

From the set of equations (\ref{eq:DS1}), we find
\bea
\label{eq:q1_all}
&& (i\slashed\partial-m_q)q_{1}(x)+Y_fG_1(x) q_{1}(x)+Y_fW_q(x,x) = 0 \,, \nonumber \\
&& \partial^2\Delta_\phi(x,y)+3\lambda\Delta_\phi(x,x)\Delta_\phi(x,y)+3\lambda G_1^2(x)\Delta_\phi(x,y) \nonumber \\
&& = Y_f\sum_{q}{\bar Q}(x,y)q_{1}(x)
+Y_f\sum_{q}{\bar q}_1(x)Q(x,y) + \delta^{(4)}(x-y) \,, \nonumber \\
&& (i\slashed\partial-m_q)S_q(x-y)+Y_fG_1(x) S_q(x-y)=\delta^{(4)}(x-y) \,, \nonumber \\  
&& \partial^2W_{q}(x,y)+3\lambda\Delta_\phi(x,x)W_{q}(x,y)+3\lambda G_1^2(x)W_{q}(x,y) =Y_f{\bar q}_1(x)S_q(x,y) \,, \nonumber \\
&& (i\slashed\partial-m_q)Q(x,y)+Y_fG_1(x)Q(x,y)+Y_f\Delta_\phi(x-y)q_{1}(x)=0 \,. 
\eea
One can use the Yukawa field propagator to write
\be
W_q(x,y)=Y_f\int d^4z\hat\Delta(x-z)\overline{q}(z)S_q(z,y) \,,
\ee
where one has
\be
\partial^2\hat\Delta(x-z)+3\lambda\Delta_\phi(x,x)\hat\Delta(x-z)+3\lambda G_1^2(x)\hat\Delta(x-z) = \delta^4(x-z) \,,
\ee
Noticing that the equation for $q_1$ has a $O(Y_f^2)$ term that, in the first approximation, can be neglected
as we assume $Y_f$ small enough for a series expansion to be meaningful.
%
Then, we have the following integral equation to be solved by iteration
%
\be
\label{eq:q1}
q_1(x)=-Y_f\int d^4y\hat{S}_q(x-y)G_1(y)q_1(x) \,, 
\ee
where
\be
(i\slashed\partial-m_q)\hat{S}_q(x-y)=\delta^{(4)}(x-y) \,,
\ee
and,
for the first iterate, we take
\be 
(i\slashed\partial-m_q)q(x)=0 \,.
\ee
In the equation for $Q(x,y)$,
the last of eq.(\ref{eq:q1_all}),
we note the ordering $O(Y_f/\lambda^\frac{1}{4})$ 
where the given ratio is assumed to be small. 
%
Then, 
%
in a first approximation and after the first iteration in eq.(\ref{eq:q1}), 
we can write
\be
\label{eq:Q}
Q(x,y)=-Y_f\int d^4z S_q(x-z)\hat{\Delta}(z-y)q(z) \,.
\ee
Furthermore,
as already stated above, $Y_f$ to be sufficiently small to grant an iteration procedure to be applied to the second of eq.(\ref{eq:q1_all}). We obtain
\bea
\Delta_\phi(x,y)&=&\hat{\Delta}(x-y)+Y_f\int d^4w\hat{\Delta}(x-w)\overline{Q}(w,y)q(w)
+Y_f\int d^4w\hat{\Delta}(x-w)\overline{q}_q(w)Q(x,w)+O(Y_f^3) \nonumber \\
&=&\hat{\Delta}(x-y)-Y_f^2\int d^4w\hat{\Delta}(x-w)\int d^4z\hat{\Delta}(z-y)\overline{q}_q(z)S_q(z-w)q(w)
\nonumber \\
&&-Y_f^2\int d^4w\hat{\Delta}(x-w)\int d^4z\hat{\Delta}(z-y)\overline{q}_q(w)S_q(w-z)q(z)+O(Y_f^3) \,.
\eea
We can Fourier-transform this equation to give 
\bea
\Delta_\phi(p)&=&\hat{\Delta}(p)-Y_f^2\int d^4w\int d^4xe^{-ipx}
\int\frac{d^4p_1}{(2\pi)^4}e^{ip_1(x-w)}\hat{\Delta}(p_1)
\int d^4z\int \frac{d^4p_2}{(2\pi)^4}e^{ip_2z}
\hat{\Delta}(p_2) \nonumber\\
&&\int\frac{d^4p_3}{(2\pi)^4}e^{ip_3z}\overline{q}_q(p_3)
\int\frac{d^4p_4}{(2\pi)^4}e^{ip_4(z-w)}S_q(p_4)
\int\frac{d^4p_5}{(2\pi)^4}e^{ip_5w}q(p_5)+c.c.+\ldots \\
&=&\hat{\Delta}(p)-
Y_f^2\hat{\Delta}(p)\int\frac{d^4p_2}{(2\pi)^4}\frac{d^4p_3}{(2\pi)^4}
\hat{\Delta}(p_2)\overline{q}_q(p_3)S_q(-p_2-p_3)q(p-p_2-p_3) 
+ {\it c.c.} + \ldots \,.
\nonumber
\eea
In order to evaluate this expression, we consider Dirac wave-packets for $q$. We can take classical wave-packets and then assume the limit of the momentum spread going to zero. This solution has been given in Ref.~\cite{Bjorken:1965sts} and can be written as follows
\be
q(x)=\int\frac{d^3p}{(2\pi)^3 \sqrt{2E}}\sum_s\left[b_{p,s}u_{p,s}e^{-ipx} +
d_{p,s}^*v_{p,s}e^{ipx}\right] \,,
\ee
provided
\bea
b_{p,s}&=&\sqrt{\frac{1}{E}}\left(\frac{1}{\sqrt{\pi}\Delta p}\right)^\frac{3}{2}e^{-\frac{1}{2}\frac{p^2}{\Delta p^2}}u_{p,s}^* \nonumber \\
d^*_{-p,s}&=&\sqrt{\frac{1}{E}}\left(\frac{1}{\sqrt{\pi}\Delta p}\right)^\frac{3}{2}e^{-\frac{1}{2}\frac{p^2}{\Delta p^2}}v_{-p,s}^* \,.
\eea
$\Delta p$ is the momentum uncertainty for the Gaussian wave-packet. The following normalization condition holds:
\be
\int d^3p\sum_s\left[|b_{p,s}|^2+|d_{-p,s}|^2\right]\stackrel{\Delta p\rightarrow 0}{=}
\int d^3p\delta^3(p)=1 \,.
\ee
Then, the Fourier transform yields
\be
q(p)=\frac{1}{\sqrt{2E}}
\sum_s\left[b_{p,s}u_{p,s}+d^*_{-p,s}v_{-p,s}\right]=
\frac{1}{\sqrt{2}E}
\left(\frac{1}{\sqrt{\pi}\Delta p}\right)^\frac{3}{2}e^{-\frac{1}{2}\frac{p^2}{\Delta p^2}}\chi \,,
\ee
where $\chi$ is a constant spinor normalized to unity and independent on momentum.
This enables us to evaluate
\bea
&&\int\frac{d^4p_2}{(2\pi)^4}\frac{d^4p_3}{(2\pi)^4}
\hat{\Delta}(p_2)\overline{q}_q(p_3)S_q(-p_2-p_3)q(p-p_2-p_3)= \nonumber \\
&&\frac{1}{2}\int\frac{d^4p_2}{(2\pi)^4}\frac{d^4p_3}{(2\pi)^4}
\hat{\Delta}(p_2)\frac{1}{E(p_3)}
\frac{1}{E(p-p_2-p_3)}
\left(\frac{1}{\sqrt{\pi}\Delta p}\right)^3e^{-\frac{1}{2}\frac{p_3^2}{\Delta p^2}}
e^{-\frac{1}{2}\frac{(p-p_2-p_3)^2}{\Delta p^2}}
\times \nonumber \\
&&{\bar\chi}\frac{-\slashed{p_2}-\slashed{p_3}-m_q}{(p_2+p_3)^2-m_q^2+i\epsilon}
\chi \,.
\eea
Taking the limit $\Delta p\rightarrow 0$, this integral can be evaluated to
\bea
&&\int\frac{d^4p_2}{(2\pi)^4}\frac{d^4p_3}{(2\pi)^4}
\hat{\Delta}(p_2)\overline{q}_q(p_3)S_q(-p_2-p_3)q(p-p_2-p_3)= \nonumber \\
&&-\hat{\Delta}(p)\int\frac{dE_2}{2\pi}
\frac{dE_3}{2\pi}\frac{(\sqrt{\pi}\Delta p)^3}{6m_q^3}=
\nonumber
-\hat{\Delta}(p)\frac{(\sqrt{\pi}\Delta p)^3}{6m_q^3}
\frac{\mu^2}{4\pi^2} \,,
\eea
that is undefined in the combined limits $\Delta p\rightarrow 0$ and $\mu\rightarrow\infty$. Now, we consider the limit of momentum localization going to zero being proportional to the quark mass as $\Delta x$ is essentially the Compton wavelength of the quark. So, we take $\sqrt{\pi}\Delta p=m_q\rightarrow 0$ and we get finally
\be
\Delta_\phi(p)=\hat{\Delta}(p) + [\hat{\Delta}(p)]^2\frac{Y_f^2\mu^2}{24\pi^2} + O(Y_f^4) \,.
\ee
Here, the effect of Yukawa interactions is due to an induced mass shift as
\be
M_n^2\rightarrow M_n^2+\frac{Y_f^2}{24\pi^2}\mu^2 \,.
\ee
This provides
\be
\lambda_n=\lambda-\frac{2K^2(i)}{3\pi^4(2n+1)^2}Y_f^2 \,.
\ee
Thus, the coupling appears to depend on the excitation the fermion interacts with.

\section{Partition function}
\label{Sect:PF}

Using Eq.~(\ref{eq:DSs}), one writes
\bea
Z[j,{\bar\eta},\eta]&=&\exp\left[iS_c[j]+\frac{i}{2}\int d^4x_1d^4x_2j(x_1)G_2(x_1-x_2)j(x_2)\right]
\times \nonumber \\
&&\int [d{\bar q}][dq]
\exp\left[i\int d^4x{\bar q}(x)\left(i{\slashed\partial}-Y_f^2\int d^4x_1G_2(x-x_1){\bar q(x_1)q(x_1)}\right)q(x)\right]\times \nonumber \\
&&\exp\left[i\int d^4x\left({\bar\eta}(x)q(x)+{\bar q}(x)\eta(x)\right)\right]+O(j^3) \,,
\eea
where $S_c[j]$ is the action evaluated at the classical solution of the scalar field. After a Fierz rearrangement, for the fermionic part we get
\bea
{\bar q}(x){\bar q(x_1)q(x_1)}q(x)&=&
\frac{1}{4}\biggl({\bar q}(x)q(x){\bar q(x_1)q(x_1)}+ \nonumber \\
&&{\bar q}(x)\gamma_\mu q(x){\bar q(x_1)\gamma^\mu q(x_1)}-
\frac{1}{8}{\bar q}(x)[\gamma^\mu,\gamma^\nu]q(x){\bar q(x_1)[\gamma_\mu,\gamma_\nu]q(x_1)}-
\nonumber \\
&&{\bar q}(x)\gamma^5q(x){\bar q(x_1)\gamma^5q(x_1)}+
{\bar q}(x)\gamma^5\gamma^\mu q(x){\bar q(x_1)\gamma^5\gamma_\mu q(x_1)} 
\biggr) \,.
\eea
We see that, in the case of a single fermionic field, we get excitations of five possible types. For now, we limit our consideration to the scalar-axial part only and perform a Stratanovich-Hubbard transformation on the fermionic part as in \cite{Hell:2008cc,Frasca:2021yuu},
by introducing two bosonic fields $\sigma$ and $\pi$ as
\bea
\exp\left[i\frac{Y_f^2}{4}\int d^4x\int d^4x_1\left({\bar q}(x)q(x)G_2(x-x_1){\bar q(x_1)q(x_1)}
-{\bar q}(x)\gamma^5q(x)G_2(x-x_1){\bar q(x_1)\gamma^5q(x_1)}\right)\right]&&
= \nonumber \\
\int[d\sigma][d\pi]
\exp\left[-i\frac{Y_f^2}{4G}\int d^4x\int d^4x_1\left(\sigma^2(x)+
\pi^2(x)\right)-i\frac{Y_f^2}{2G}
\int d^4xd^4y{\bar q}(x)G_2(x-y)\left(\sigma\left(\frac{x+y}{2}\right)+\gamma^5\pi\left(\frac{x+y}{2}\right)\right)q(y)\right] \,, &&
\eea
%
The local limit is then obtained by taking $G_2(x_2-x_1) \propto \delta^4(x_2-x_1)$ starting from Eq.~(\ref{eq:G2}) at zero momenta $p\rightarrow 0$. 

Finally, we can integrate out the fermionic degrees of freedom to get the effective Lagrangian
\be
\mathcal L_{\rm eff}=-\frac{1}{2G}(\sigma^2+\pi^2)-i\operatorname{tr}\ln({i\slashed\partial-m_q-Y_f\sigma-Y_f\gamma^5\pi}) \,.
\ee
This provides a definition for $G$ as
\be
\label{eq:GNJL}
G=2Y_f^2\Delta_\phi(0)=\frac{Y_f^2}{m_0^2}\sum_{n=0}^\infty\frac{B_n}{(2n+1)^2(\pi/2K(-1))^2}\propto \frac{Y_f^2}{m_0^2} \,, 
\ee
where $m_0$ is found is Eq.(\ref{eq:m0}), and the coefficients in the series are given by
\begin{equation}
   B_n=(2n+1)^2\frac{\pi^3}{4K^3(-1)}\frac{e^{-(n+\frac{1}{2})\pi}}{1+e^{-(2n+1)\pi}} \,.
\end{equation}
See below for an extended derivation of the NJL constant.

The fermionic determinant can be expanded noticing that
\be
\operatorname{tr}\ln({i\slashed\partial-m_q-Y_f\sigma-Y_f\gamma^5\pi})=
\operatorname{tr}\ln({i\slashed\partial-m_q})+
\operatorname{tr}\ln(1-(i\slashed\partial-m_q)^{-1}(Y_f\sigma+Y_f\gamma^5\pi)) \,.
\ee
One can neglect the first term (a constant) and expand the second logarithm to obtain
\bea
&&\operatorname{tr}\ln(1-Y_f(i\slashed\partial-m_q)^{-1}(\sigma+\gamma^5\pi))= \nonumber \\
&&\operatorname{tr}\left(-(i\slashed\partial-m_q)^{-1}(Y_f\sigma+Y_f\gamma^5\pi)
-\frac{1}{2}(i\slashed\partial-m_q)^{-2}(Y_f^2\sigma^2+Y_f^2\pi^2+2Y_f^2\gamma^5\sigma\pi)+\ldots
\right) \,.
\eea
This will yield the effective Lagrangian \cite{Ebert:1982pk} 
\bea
\label{eq:Leff}
\mathcal L_{\rm eff}&=&-\frac{1}{2G}(\sigma^2+\pi^2)+\frac{1}{2}(8Y_f^2I_1)(\sigma^2+\pi^2)+\frac{1}{2}(4Y_f^2I_2)[(\partial\sigma)^2+(\partial\pi)^2] \\
&&-\frac{1}{2}4m_q^2(4Y_f^2I_2)\sigma^2
-8Y_f^3(8m_qI_2)\sigma(\sigma^2+\pi^2)-2Y_f^4I_2(\sigma^2+\pi^2)^2 + \ldots \,,
\nonumber
\eea
where
\bea
I_1(m)&=&i\int^\Lambda\frac{d^4p}{(2\pi)^4}\frac{1}{p^2-m^2} \nonumber \\
I_2(m)&=&-i\int^\Lambda\frac{d^4p}{(2\pi)^4}\frac{1}{(p^2-m^2)^2} \,.
\eea
These are divergent integrals to be regularized e.g.~by a cut-off $\Lambda$. By requiring
\be
\left.\frac{\delta L_{eff}}{\delta\sigma}\right|_{\sigma = m,\pi = 0}=0 \,,
\ee
one straightforwardly arrives at the gap equation 
\be
m=m_q+8mGI_1(m)
\ee
for the fermion mass.
%
Then, we introduce the renormalized fields
\be
\sigma\rightarrow (4Y_f^2I_2)^\frac{1}{2}\sigma \qquad \pi\rightarrow (4Y_f^2I_2)^\frac{1}{2}\pi \,,
\ee
and we get the final result, after chiral symmetry breaking,
\bea
\label{eq:Leff2}
L&=&\frac{1}{2}[(\partial\sigma)^2+(\partial\pi)^2]-\frac{1}{2}(4m^2+m_q^2)\sigma^2 -\frac{1}{2}\frac{m_qm}{I_2G}\pi^2 \nonumber \\
&&
-\frac{m}{I_2^\frac{1}{2}}\sigma(\sigma^2+\pi^2)-\frac{1}{8I_2}(\sigma^2+\pi^2)^2+\ldots \,.
\eea
Such a model is manifestly renormalizable as expected.

\section{Fermion gap equation and scalar self-coupling}
\label{Sect:GE}

\subsection{Local case}

Let us explore the properties of the Yukawa model considering first 
its local limit 
\be
\Delta(x-y)\propto \delta^4(x-y) \,.
\ee
In this case, we can get an implicit function for the fermion mass and the scalar self-coupling. Indeed, we have \cite{Klevansky:1992qe}
\be
M = m_q+\frac{1}{2\pi^2}\frac{M}{m_0^2+\frac{1}{G}}\left[\Lambda^2-M^2\ln\left(1+\frac{\Lambda^2}{M^2}\right)\right] \,,
\ee
where $m_0$ is the classical mass gap. The value of $G$ is given by (see. eq.~(\ref{eq:GNJL}))
\be
G=\kappa\frac{Y_f^2}{\mu^2(\lambda/2)^\frac{1}{2}} \,,
\ee
with $\kappa$ being a proportionality constant of the order of unity, and
\be
\label{eq:m0}
m_0=\frac{\pi}{2K(i)}\left(\frac{\lambda}{2}\right)^\frac{1}{4}\mu \,.
\ee
This yields
\be
    \sqrt{\lambda} = \frac{M}{\sqrt{2}\pi^2(M-m_q)}\frac{1}{\frac{\pi^2}{4K^2(i)}+\frac{1}{\kappa Y_f^2}}\left[\frac{\Lambda^2}{\mu^2}-\frac{M^2}{\mu^2}\ln\left(1+\frac{\Lambda^2}{M^2}\right)\right] \,.
\ee

\begin{figure}[H]
\centering
\includegraphics[width=\textwidth]{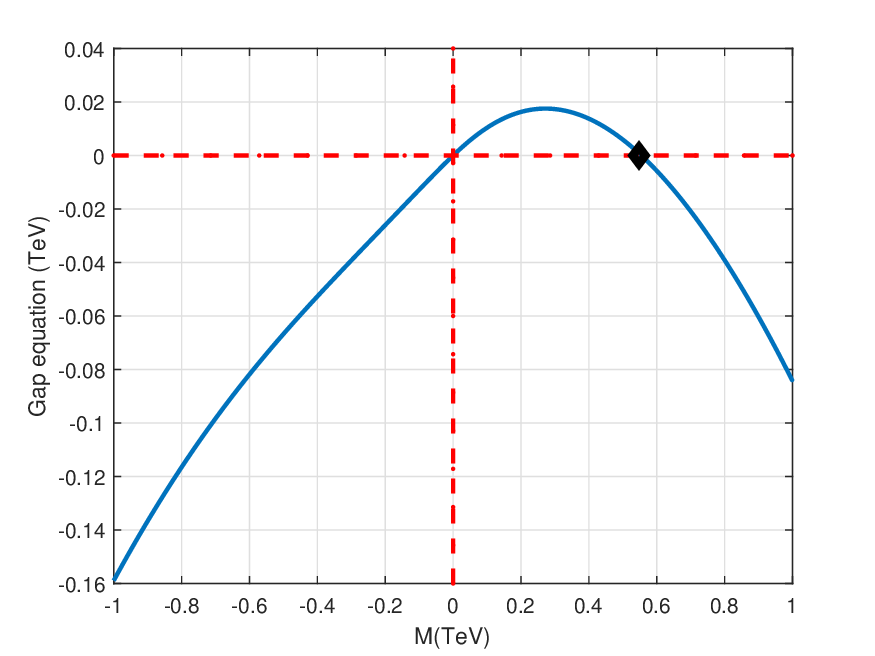}
\caption{\it The chiral condensate given by an effective mass for the fermion when the curve crosses the abscissa. The trivial solution is also present. We have assumed $m_q=0$ for the sake of simplicity as this choice does not change the conclusions. The energy scale is arbitrary and taken in the TeV range.
}
\label{fig1}
\end{figure}

To summarise, we ended up effectively having two energy scales $\mu$ and $\Lambda$. While the former is an integration constant of the scalar field, the latter is the regularising cut-off to yield UV-finite integrals in the NJL model. A physically consistent choice is to set them equal to each other. Such a theory is not confining as a meson state is permitted to decay into two free bare fermions. The only condensate in the theory is the chiral one as shown in Fig.~\ref{fig1}.

\subsection{Non-local case}

We consider the propagator for the scalar sector as proposed in Refs.~\cite{Frasca:2013kka,Frasca:2011bd} (see Eq.~(\ref{eq:G2}) above). Then, the gap equation can be written down in the form \cite{Frasca:2011bd,Frasca:2021yuu,Frasca:2022pjf}
\be
\label{eq:mg1}
M=m_q+{\cal C}(p)v \,,
\ee
where
\be
\label{eq:mg2}
v = \frac{4}{m_0^2+\frac{1}{G}}\int\frac{d^4p}{(2\pi)^4}{\cal C}(p)\frac{M(p)}{p^2+M^2(p)} \,,
\ee
for a single flavor, and $m_0$ given by Eq.~(\ref{eq:ms}). In our case, the kernel is found to be
\begin{equation}
\label{eq:GG}
   \frac{G}{2}{\cal C}(p)={\cal G}(p)=-\frac{1}{2}Y_f^2\sum_{n=0}^\infty\frac{B_n}{p^2-(2n+1)^2m_0^2+i\epsilon} \,,
\end{equation}
with $G$ being the NJL constant that in our case is given by Eq.~(\ref{eq:GNJL}).

In the general case, we can write for the Yukawa sector (in the Euclidean limit)
\begin{equation}
\Delta(p)=\sum_{n=0}^\infty\frac{B_n}{p^2+m_n^2}
\end{equation}
with the mass spectrum, $m_n$, given in Eq.~(\ref{eq:ms}) at the leading order. So, we have to evaluate
\begin{equation}
M=m_q+\frac{Y_f^2}{2}\int\frac{d^4p}{(2\pi)^4}\sum_{n=0}^\infty\frac{B_n}{p^2+m_n^2}\frac{M}{p^2+M^2} \,.
\end{equation}
Here, the integral can be computed exactly when a cut-off $\Lambda$ is used, as usual for NJL models. This yields
\begin{equation}
\label{eq:gape}
M=m_q+\frac{Y_f^2}{16\pi^2}\sum_{n=0}^\infty\frac{B_nM}{2(m_n^2-M^2)}
\left[m_n^2\ln\left(1+\frac{\Lambda^2}{m_n^2}\right)-M^2\ln\left(1+\frac{\Lambda^2}{M^2}\right)\right] \,.
\end{equation}

The idea to understand the fermion confinement is strongly linked to the expected
behavior of the roots of the gap equation~(\ref{eq:gape}), i.e.\ the poles of the
fermion propagator. The idea presented here is identical with the one presented
in Ref.~\cite{Rezaeian:2004nf}, though without employing a general model for non-locality. To represent the physical propagating degrees of freedom, for
these poles one should expect solutions on the real axis. The effect of the
gluonic interaction is to move such poles in the complex plane so that no
decay into such degrees of freedom is ever expected, and hence the fundamental fermions in the theory never propagate freely. Eq.~(\ref{eq:gape}) is amenable to a numerical treatment. It should be solved with the conditions $M\ge 0$ and $M\ll\Lambda$ that is, the effective mass of the fermion should not exceed the UV cut-off representing, at least, the boundary of the region where the high-energy regime starts to set in (generally taken at $\Lambda\approx 1\ {\rm GeV}$ for similarity with QCD). We can normalize this equation to the cut-off $\Lambda$ by introducing the new variables $u_1=m_0/\Lambda$, $u_2=p/\Lambda$ and $w=M/\Lambda$ having taken $m_n=(2n+1)m_0$.  The mass $m_0$ can be assumed to be, for instance, that of the $\sigma$ 
meson or $f(500)$ that we fix to $m_0=0.417\ {\rm GeV}$ to keep on utilizing QCD physics as a suitable analogue. Then,
\begin{equation}
\label{eq:Meff}
w=\frac{m_q}{\Lambda}+\kappa\alpha_s\sum_{n=0}^\infty\frac{B_ny}{(2n+1)^2u_1^2-u_2^2}\left[(2n+1)^2u_1^2\ln\left(1+\frac{1}{(2n+1)^2u_1^2}\right)-u_2^2\ln\left(1+\frac{1}{u_2^2}\right)\right] \,,
\end{equation}
with $\kappa=1/8\pi^2$ and $\alpha_s=Y_f^2/4\pi$, exploiting the analogy with strong interactions. We note that the cut-off dependence is effectively present in the ratios, while the ratio $m_q/\Lambda$ is negligibly small for light fermions and can be omitted. 

\begin{figure}[H]
\centering
\includegraphics[width=\textwidth]{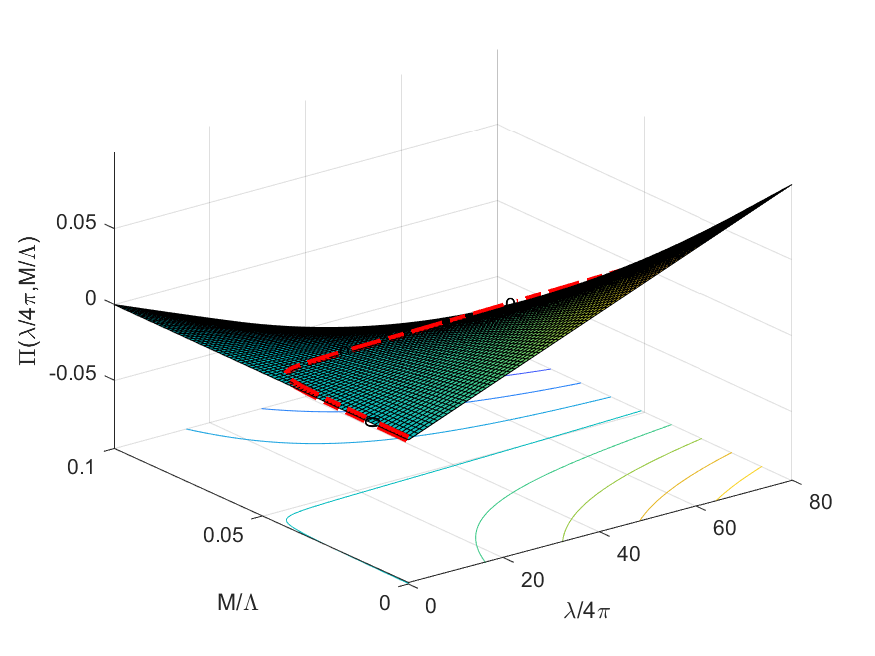}
\caption{\it The gap equation $\Pi(\alpha_s,M/\Lambda)$ as a function of the scalar self-coupling $\alpha_s=Y_f^2/4\pi$, and the effective fermion mass $M$. The red 
dashed
line is where zeros lie granting a solution to the gap equation.
}
\label{fig:Meff}
\end{figure}

The result for Eq.~(\ref{eq:Meff}), written in implicit form when the rhs of the equation is moved on the lhs, reads
\begin{equation}
\label{eq:mu}
\Pi(u_1,u_2)=w-\frac{m_q}{\Lambda}-\kappa\alpha_s\sum_{n=0}^\infty\frac{B_nu_2}{(2n+1)^2u_1^2-u_2^2}\left[(2n+1)^2u_1^2\ln\left(1+\frac{1}{(2n+1)^2u_1^2}\right)-Y_f^2\ln\left(1+\frac{1}{u_2^2}\right)\right] \,,
\end{equation}
and is illustrated in Fig.~\ref{fig:Meff}. The zeros of the function $\Pi(u_1,u_2)$ are of interest here. We have to solve Eqs.~(\ref{eq:mg1}) and (\ref{eq:mg2}) numerically and evaluate the fermion mass as a function of $\lambda$ -- the self-coupling of the scalar sector. 
%
The numerical result for the zeros is represented by red 
dashed
line in Fig.~\ref{fig:Meff}.

It is worth noticing that Eq.~(\ref{eq:mg1}) can give rise to different condensates as the coupling increases \cite{Frasca:2022lwp}. The fermion propagator admits a pole on the real axis when a chiral condensate forms. With a further increase of the coupling, this pole becomes complex and, quite possibly, a gluon condensate forms. The bare fermion is no longer a true degree of freedom of the theory in this regime, entering a confining phase where only bounded states of the fermion field are found as asymptotic states. Hence, the triviality of the scalar field, in the presence of Yukawa interactions, yields non-trivial non-perturbative physics in the IR regime of the theory, in particular, indicating the signatures of confinement.

\section{Discussion and conclusions}
\label{Sect:CC}

We have shown that, using this given exact solution of the set of Dyson-Schwinger equations for a quartic scalar field, surely not unique, interaction with a fermion could entail confinement. Therefore, the technology is very similar to that discussed in a full QCD framework \cite{Frasca:2021yuu,Frasca:2022lwp,Frasca:2022pjf}. The relevant results are that, also for the Yukawa model, a non-local Nambu-Jona-Lasinio model comes out with the property of being confining and only bound states of the fermion fields could be observed in principle. From this analysis, it appears that confinement is a rather ubiquitous effect for several quantum field theories for some possible choices of the vacuum state. The kind of physical scenario that is obtained is that of a chiral condensate of fermions transiting to an instanton liquid for the scalar sector emulating the gluon field of standard QCD. 


Some important conclusions could be drawn from our analysis. We were able to get an exact beta function for the scalar sector represented by a massless $\phi^4$ theory. Our RG study confirms that the scalar theory, with no further interactions, 
behaves as generally expected in agreement with studies on triviality. About the universality class, a recent work \cite{Lundow:2022ltb} casts a shadow about the scalar theory. We hope to discuss this point in our future studies.

An interesting aspect of our finding is that quantum effects generate a mass term in the Dyson-Schwinger set of equations and the spectrum of the theory is that of a harmonic oscillator. The propagator displays such poles, in agreement with 
K\"all\'en--Lehmann
representation, but no bound states appear as it should be for a trivial non-interacting theory. The choice of the ground state as a Fubini instanton provides a leading term in the beta function somewhat different from weak perturbation theory expectations while the next-to-leading order term, in the same limit, is fully recovered in our computation, granting a proper UV-limit to the theory.

The introduction of a fermion with a Yukawa coupling entails some interesting consequences. We get an effect on the mass spectrum of the theory due to such Yukawa coupling. The fermion appears to shift the mass spectrum of the scalar theory and this could be formally interpreted as a change in the self-coupling of the scalar theory depending on the excitations of the field.

When the scalar degree of freedom is integrated out, a non-local Nambu-Jona-Lasinio model is obtained for the fermion field. Differently, from a local NJL, in this case we can have confinement. We refer to such a behavior when the fermion exits from the physical spectrum of the theory. Indeed, in the low-energy limit, we recover an energy range where this happens. Going to higher energies, we get an effective mass for the fermion instead showing a breaking of the initial chiral symmetry of the theory, exactly in the same way as for QCD. So, in our presumed confined phase, we observe a transition from a chiral condensate to an instanton condensate of the scalar field that hinders a possible physical state for the fermion as an asymptotic state of the theory.

The analysis we presented here should find large applications in BSM physics and cosmological settings where an analytical treatment of the Yukawa theory is missing. 
In the framework of the Standard Model instead, weak perturbation theory works really well and we should expect that deviations like the one we discussed in this paper could appear at energy so high that the physics could change significantly. We hope to address some of this matter in the future.



\medskip

\section{Acknowledgements}
\label{Asck}

Authors acknowledge Ignacy Nalecz for reading and suggestion on the manuscript. Authors thank Roman Pasechnik and Zhi-Wei Wang for collaborating during the early stages of the work and helpful insights regarding the work.

\newpage

\section*{Appendix}

In this appendix we show the behavior of $G_3$ and $G_4$ correlation functions. Indeed, the corresponding Dyson-Schwinger equations take the form (we assume $G_2$ translation invariant as discussed in the main text) \cite{Frasca:2022kfy}
\begin{eqnarray}
    &&\partial^2G_3(x,y,z)+\lambda\left[6G_1(x)G_2(x-y)G_2(x-z)+3G_1^2(x)G_3(x,y,z)\right. \\ \nonumber
    &&+3G_2(x-z)G_3(x,x,y)+3G_2(x-y)G_3(x,x,z) \\ \nonumber
    &&\left.+3G_2(0)G_3(x,y,z)+3G_1(x)G_4(x,x,y,z)+G_5(x,x,x,y,z)\right]=0 \\ \nonumber
    &&\\ \nonumber
    &&\partial^2G_4(x.y,z,w)+\lambda\left[6G_2(x-y)G_2(x-z)G_2(x-w)
    \right. \\ \nonumber
    &&+6G_1(x)G_2(x-y)G_3(x,z,w)+6G_1(x)G_2(x-z)G_3(x,y,w)\\ \nonumber
    &&+6G_1(x)G_2(x-w)G_3(x,y,z)+3G_1^2(x)G_4(x,y,z,w) \\ \nonumber
    &&+3G_2(x-y)G_4(x,x,z,w)+3G_2(x-z)G_4(x,x,y,w)  \\ \nonumber
    &&+3G_2(x-w)G_4(x,x,y,z)+3G_2(0)G_4(x,y,z,w) \\ \nonumber
    &&\left.+3G_1(x)G_5(x,x,y,z,w)+G_6(x,x,x,y,z,w)\right]=0 \\ \nonumber
    &\vdots&
\end{eqnarray}
that are solved by
\begin{equation}
\label{eq:G_3}
   G_3(x,y,z)=-6\lambda\int dx_1 G_2(x-x_1)G_1(x_1)G_2(x_1-y)G_2(x_1-z) 
\end{equation}
and it is easy to verify that $G_3(x,x,z)=G_3(x,y,x)=0$ using the property of Heaviside function $\theta(x)\theta(-x)=0$, and
\begin{eqnarray}
    &&G_4(x,y,z,w)=-6\lambda\int dx_1 G_2(x-x_1)G_2(x_1-y)G_2(x_1-z)G_2(x_1-w) \\ \nonumber
    &&-6\lambda\int dx_1G_2(x-x_1)\left[G_1(x_1)G_2(x_1-y)G_3(x_1,z,w)\right. \\ \nonumber
    &&\left.+G_1(x_1)G_2(x_1-z)G_3(x_1,y,w)
    +G_1(x_1)G_2(x_1-w)G_3(x_1,y,z)\right].
\end{eqnarray}
and it is not difficult to verify that $G_4(x,x,x,y)=0$. From a standpoint of functional calculus, this implies that our solution is just given by the following functional Taylor series
\bea
\label{eq:DSs}
\phi(x)=\sum_{i=1}^\infty\int G_i(x_1,\ldots,x_i)\left[\prod_{k=2}^ij(x_k) d^4x_k\right].
\eea
This definition implies the absence of an integration for $i=1$
The index $i$ gives also the number of the independent variables of the correlation functions $G_i$.


The theory is indeed stable. This can be seen with the following argument. The Hamiltonian of the system is given by
\begin{equation}
    H=\int d^3x\left[\frac{1}{2}(\partial_t\phi)^2+\frac{1}{2}(\nabla\phi)^2+\frac{\lambda}{4}\phi^4\right].
\end{equation} 
We linearize it around the classical solution (given by $G_1$ in (\ref{solG1})) 
\begin{equation}
    \phi(x)=\phi_0(x)+\delta\phi(x)\,,
\end{equation}
yielding
\begin{equation}
   H=H_0+\int d^3x\left[\frac{1}{2}(\partial_t\delta\phi)^2+\frac{1}{2}(\nabla\delta\phi)^2
   +\frac{3}{2}\lambda\phi_0^2\delta\phi^2\right]+O\left(\delta\phi^3\right),
\end{equation}
being $H_0$ the contribution coming from the classical solution. The linear part can be diagonalized with a Fourier series provided we are able to get the eigenvalues and the eigenvectors of the operator
\begin{equation}
   W_0=-\Box+3\lambda\phi_0^2(x).
\end{equation}
It is not difficult to realize that there is a zero mode. We give the solutions for both the zero and non-zero modes. The spectrum is continuous with eigenvalues 0 and $3\mu^2\sqrt{\lambda/2}$ with $\mu$ varying continuously from 0 to infinity. The zero-mode solution has the aspect
\begin{equation}
  \omega_0(x,\mu)=a_0\,{\rm cn}(p\cdot x+\theta,i)\,{\rm dn}(p\cdot x+\theta,i)
\end{equation}
being $a_0$ a normalization constant. Non-zero modes are given by
\begin{equation}
  \omega(x,\mu)=a'\,{\rm sn}(p\cdot x+\theta,i)\,{\rm dn}(p\cdot x+\theta,i).
\end{equation}
with $a'$ again a normalization constant. These hold on-shell, that is when $p^2=\mu^2\sqrt{\lambda/2}$. Since the spectrum is continuous, these eigenfunctions are not normalizable. Therefore, we note that there is a doubly degenerate set of zero modes spontaneously breaking translational invariance and the $Z_2$ symmetry of the theory. 
For a given $\mu$ parameter, $Z_2$ symmetry is spontaneously broken through this zero mode. This mode disappears when $\mu=0$, as it should, and one goes back to a standard textbook solution.


\begin{thebibliography}{99}
\bibitem{Workman:2022ynf}
R.~L.~Workman \textit{et al.} [Particle Data Group],
PTEP \textbf{2022}, 083C01 (2022)
doi:10.1093/ptep/ptac097

\bibitem{ATLAS:2022vkf}
ATLAS Collaboration,
Nature \textbf{607}, no.7917, 52-59 (2022)
[erratum: Nature \textbf{612}, no.7941, E24 (2022)]
doi:10.1038/s41586-022-04893-w
[arXiv:2207.00092 [hep-ex]].

\bibitem{ATLAS:2022jtk}
ATLAS Collaboration,
[arXiv:2211.01216 [hep-ex]].

\bibitem{Aizenman:2019yuo}
M.~Aizenman and H.~Duminil-Copin,
Annals Math. \textbf{194}, no.1, 163 (2021)
doi:10.4007/annals.2021.194.1.3
[arXiv:1912.07973 [math-ph]].

\bibitem{Dyson1949} F. J. Dyson, Phys. Rev. 75, 486 (1949).

\bibitem{Dyson1952} F. J. Dyson, Phys. Rev. 85, 631 (1952).

\bibitem{Landau1955} L. D. Landau, {\it Niels Bohr and the Development of Physics}, London, Pergamon Press LTD, p. 52, 1955.

\bibitem{LandauPomeranchuk1955} L. D. Landau and I. Ya Pomeranchuk, Dokl. Akad. Nauk SSSR {\bf 102}, 489 (1955).

\bibitem{Landau1956} L. D. Landau, A. Abrikosov and L. Halatnikov, Suppplemento AL, {\bf 111}, Il Nuovo Cimento (1956).

\bibitem{Callaway1986} D. J. E. Callaway and R. Petronzio, Nucl. Phys. B{\bf 277}, 50 (1986).

\bibitem{Callaway1988} D. J. E. Callaway, Physics Reports {\bf 167}, 241 (1998).

\bibitem{Wilson} K. G. Wilson, Rev. Mod. Phys. {\bf 47}, 773 (1975).

\bibitem{Mann:2017wzh}
R.~Mann, J.~Meffe, F.~Sannino, T.~Steele, Z.~W.~Wang and C.~Zhang,
Phys. Rev. Lett. \textbf{119} (2017) 261802.

\bibitem{Antipin:2018zdg}
O.~Antipin, N.~A.~Dondi, F.~Sannino, A.~E.~Thomsen and Z.~W.~Wang,
Phys. Rev. D \textbf{98} (2018) 016003.

\bibitem{Molinaro:2018kjz}
E.~Molinaro, F.~Sannino and Z.~W.~Wang,
Phys. Rev. D \textbf{98} (2018) 115007.

\bibitem{Wang:2018yer}
Z.~W.~Wang, A.~Al Balushi, R.~Mann and H.~M.~Jiang,
Phys. Rev. D \textbf{99} (2019) 115017.

\bibitem{Sannino:2019sch}
F.~Sannino, J.~Smirnov and Z.~W.~Wang,
Phys. Rev. D \textbf{100} (2019) 075009.

\bibitem{Cacciapaglia:2023ghp}
G.~Cacciapaglia, A.~Deandrea, R.~Pasechnik and Z.~W.~Wang,
[arXiv:2302.11671 [hep-th]].

\bibitem{Cacciapaglia:2020qky}
G.~Cacciapaglia, A.~S.~Cornell, C.~Cot and A.~Deandrea,
Phys. Rev. D \textbf{104} (2021) 075012.

\bibitem{Christiansen:2017gtg}
N.~Christiansen and A.~Eichhorn,
Phys. Lett. B \textbf{770} (2017) 154-160.

\bibitem{Callan1970} C. Callan, Phys. Rev. D2, 1541 (1970).

\bibitem{Symanzik1970} K. Symanzik, Comm. Math. Phys. 18, 227 (1970).

\bibitem{Symanzik1971} K. Symanzik, Comm. Math. Phys. 23, 49 (1971).

\bibitem{Peskin1995} M. E. Peskin and D. V. Schroeder, {\it an Introxuction to Quantum Field Theory}, Addison-Wesley, Reading 1995, 2nd edition,
Westview Press 2015.

\bibitem{Wilson:1973jj}
K.~G.~Wilson and J.~B.~Kogut,
Phys. Rept. \textbf{12}, 75-199 (1974)
doi:10.1016/0370-1573(74)90023-4

\bibitem{Luscher:1987ek}
M.~Luscher and P.~Weisz,
Nucl. Phys. B \textbf{295}, 65-92 (1988)
doi:10.1016/0550-3213(88)90228-3

\bibitem{Hasenfratz:1987eh}
A.~Hasenfratz, K.~Jansen, C.~B.~Lang, T.~Neuhaus and H.~Yoneyama,
Phys. Lett. B \textbf{199}, 531-535 (1987)
doi:10.1016/0370-2693(87)91622-4

\bibitem{Heller:1992js}
U.~M.~Heller, H.~Neuberger and P.~M.~Vranas,
Nucl. Phys. B \textbf{399}, 271-348 (1993)
doi:10.1016/0550-3213(93)90499-F
[arXiv:hep-lat/9207024 [hep-lat]].

\bibitem{Callaway:1988ya}
D.~J.~E.~Callaway,
Phys. Rept. \textbf{167}, 241 (1988)
doi:10.1016/0370-1573(88)90008-7

\bibitem{Rosten:2008ts}
O.~J.~Rosten,
JHEP \textbf{07}, 019 (2009)
doi:10.1088/1126-6708/2009/07/019
[arXiv:0808.0082 [hep-th]].

\bibitem{Gell-Mann:1954yli}
M.~Gell-Mann and F.~E.~Low,
Phys. Rev. \textbf{95}, 1300-1312 (1954)
doi:10.1103/PhysRev.95.1300

\bibitem{Gockeler:1997dn}
M.~Gockeler, R.~Horsley, V.~Linke, P.~E.~L.~Rakow, G.~Schierholz and H.~Stuben,
Phys. Rev. Lett. \textbf{80}, 4119-4122 (1998)
doi:10.1103/PhysRevLett.80.4119
[arXiv:hep-th/9712244 [hep-th]].

\bibitem{Gies:2004hy}
H.~Gies and J.~Jaeckel,
Phys. Rev. Lett. \textbf{93}, 110405 (2004)
doi:10.1103/PhysRevLett.93.110405
[arXiv:hep-ph/0405183 [hep-ph]].

\bibitem{Cabibbo:1979ay}
N.~Cabibbo, L.~Maiani, G.~Parisi and R.~Petronzio,
Nucl. Phys. B \textbf{158}, 295-305 (1979)
doi:10.1016/0550-3213(79)90167-6

\bibitem{Kuti:1987nr}
J.~Kuti, L.~Lin and Y.~Shen,
Phys. Rev. Lett. \textbf{61}, 678 (1988)
doi:10.1103/PhysRevLett.61.678

\bibitem{Hambye:1996wb}
T.~Hambye and K.~Riesselmann,
Phys. Rev. D \textbf{55}, 7255-7262 (1997)
doi:10.1103/PhysRevD.55.7255
[arXiv:hep-ph/9610272 [hep-ph]].

\bibitem{Fodor:2007fn}
Z.~Fodor, K.~Holland, J.~Kuti, D.~Nogradi and C.~Schroeder,
PoS \textbf{LATTICE2007}, 056 (2007)
doi:10.22323/1.042.0056
[arXiv:0710.3151 [hep-lat]].

\bibitem{Gerhold:2008mb}
P.~Gerhold, K.~Jansen and J.~Kallarackal,
PoS \textbf{LATTICE2008}, 067 (2008)
doi:10.22323/1.066.0067
[arXiv:0810.4447 [hep-lat]].

\bibitem{Yndurain:1991vm}
F.~J.~Yndurain,
in R. Akhoury,  B. De Wit, P. van Nieuwenhuizen, H. Veltman, ``Gauge theories - past and future: In commemoration of the 60th birthday of M. Veltman - Ann Arbor, Michigan, 16 - 18 May 1991''
(World Scientific Publishing, Singapore, 1992), pp. 337-353.

\bibitem{Pasechnik:2021ncb}
R.~Pasechnik and M.~\v{S}umbera,
Universe \textbf{7}, no.9, 330 (2021)
doi:10.3390/universe7090330
[arXiv:2109.07600 [hep-ph]].

\bibitem{Nambu:1961tp}
  Y.~Nambu and G.~Jona-Lasinio,
  ``Dynamical Model of Elementary Particles
  Based on an Analogy with Superconductivity. 1.,''
  Phys.\ Rev.\ \textbf{122}, 345-358 (1961).

\bibitem{Nambu:1961fr}
  Y.~Nambu and G.~Jona-Lasinio,
  ``Dynamical Model of Elementary Particles
  Based on an Analogy with Superconductivity. 2.,''
  Phys.\ Rev.\ \textbf{124}, 246-254 (1961).

\bibitem{Klevansky:1992qe}
  S.~P.~Klevansky,
  ``The Nambu-Jona-Lasinio model of quantum chromodynamics,''
  Rev.\ Mod.\ Phys.\ \textbf{64}, 649 (1992).

\bibitem{GomezDumm:2006vz}
  D.~Gomez Dumm, A.~G.~Grunfeld and N.~N.~Scoccola,
  ``On covariant nonlocal chiral quark models with separable interactions,''
  Phys.\ Rev.\ D \textbf{74}, 054026 (2006).

\bibitem{Hell:2008cc}
  T.~Hell, S.~R\"ossner, M.~Cristoforetti, W.~Weise,
  ``Dynamics and thermodynamics of a non-local PNJL model
  with running coupling,''
  Phys.\ Rev.\  {\bf D79} (2009) 014022.


\bibitem{Hell:2008cc}
T.~Hell, S.~Roessner, M.~Cristoforetti and W.~Weise,
Phys. Rev. D \textbf{79}, 014022 (2009)
doi:10.1103/PhysRevD.79.014022
[arXiv:0810.1099 [hep-ph]].

\bibitem{Frasca:2021yuu}
M.~Frasca, A.~Ghoshal and S.~Groote,
Phys. Rev. D \textbf{104}, no.11, 114036 (2021)
doi:10.1103/PhysRevD.104.114036
[arXiv:2109.05041 [hep-ph]].

\bibitem{Bowler:1994ir}
  R.~D.~Bowler and M.~C.~Birse,
  ``A Nonlocal, covariant generalization of the NJL model,''
  Nucl.\ Phys.\ A \textbf{582}, 655 (1995).

\bibitem{Roberts:1994dr}
  C.~D.~Roberts and A.~G.~Williams,
  ``Dyson-Schwinger equations and their application to hadronic physics,''
  Prog.\ Part.\ Nucl.\ Phys.\ \textbf{33}, 477-575 (1994).

\bibitem{Bender:1999ek}
C.~M.~Bender, K.~A.~Milton and V.~Savage,
Phys. Rev. D \textbf{62}, 085001 (2000)
doi:10.1103/PhysRevD.62.085001
[arXiv:hep-th/9907045 [hep-th]].

\bibitem{Frasca:2015yva}
M.~Frasca,
Eur. Phys. J. Plus \textbf{132}, no.1, 38 (2017)
[erratum: Eur. Phys. J. Plus \textbf{132}, no.5, 242 (2017)]
doi:10.1140/epjp/i2017-11321-4
[arXiv:1509.05292 [math-ph]].


\bibitem{Frasca:2021mhi}
M.~Frasca, A.~Ghoshal and S.~Groote,
Nucl. Part. Phys. Proc. \textbf{318-323}, 138-141 (2022)
doi:10.1016/j.nuclphysbps.2022.09.029
[arXiv:2109.06465 [hep-ph]].

\bibitem{Wetterich:1992yh}
C.~Wetterich,
Phys. Lett. B \textbf{301}, 90-94 (1993)
doi:10.1016/0370-2693(93)90726-X
[arXiv:1710.05815 [hep-th]].


\bibitem{Frasca:2022kfy}
M.~Frasca, A.~Ghoshal and N.~Okada,
J. Phys. G \textbf{51}, no.3, 035001 (2024)
doi:10.1088/1361-6471/ad170e
[arXiv:2201.12267 [hep-th]].

\bibitem{Calcagni:2022tls}
G.~Calcagni, M.~Frasca and A.~Ghoshal,
[arXiv:2206.09965 [hep-th]].


\bibitem{Calcagni:2022gac}
G.~Calcagni, M.~Frasca and A.~Ghoshal,
Int. J. Mod. Phys. D \textbf{33}, no.01, 2350111 (2024)
doi:10.1142/S0218271823501110
[arXiv:2211.06957 [hep-th]].

\bibitem{Wipf:2021mns}
A.~Wipf,
``Statistical Approach to Quantum Field Theory: An Introduction,''
Lect. Notes Phys. \textbf{992}, 1-554 (2021).
ISBN 978-3-030-83262-9, 978-3-030-83263-6
doi:10.1007/978-3-030-83263-6

\bibitem{Peskin:1995ev}
M.~E.~Peskin and D.~V.~Schroeder,
``An Introduction to quantum field theory,''
Addison-Wesley, 1995,
ISBN 978-0-201-50397-5. Page 404.


\bibitem{Frasca:2022lwp}
M.~Frasca, A.~Ghoshal and S.~Groote,
Phys. Lett. B \textbf{846}, 138209 (2023)
doi:10.1016/j.physletb.2023.138209
[arXiv:2202.14023 [hep-ph]].

\bibitem{Frasca:2022pjf}
M.~Frasca, A.~Ghoshal and S.~Groote,
Nucl. Part. Phys. Proc. \textbf{324-329}, 85-89 (2023)
doi:10.1016/j.nuclphysbps.2023.01.019
[arXiv:2210.02701 [hep-ph]].

\bibitem{Chaichian:2018cyv}
M.~Chaichian and M.~Frasca,
Phys. Lett. B \textbf{781}, 33-39 (2018)
doi:10.1016/j.physletb.2018.03.067
[arXiv:1801.09873 [hep-th]].

\bibitem{Frasca:2015wva}
M.~Frasca,
Eur. Phys. J. Plus \textbf{131}, no.6, 199 (2016)
doi:10.1140/epjp/i2016-16199-x
[arXiv:1504.02299 [hep-ph]].

\bibitem{Frasca:2019ysi}
M.~Frasca,
Eur. Phys. J. C \textbf{80}, no.8, 707 (2020)
doi:10.1140/epjc/s10052-020-8261-7
[arXiv:1901.08124 [hep-ph]].

\bibitem{Frasca:2017slg}
M.~Frasca,
Nucl. Part. Phys. Proc. \textbf{294-296}, 124-128 (2018)
doi:10.1016/j.nuclphysbps.2018.02.005
[arXiv:1708.06184 [hep-ph]].

\bibitem{Frasca:2016sky}
M.~Frasca,
Eur. Phys. J. C \textbf{77}, no.4, 255 (2017)
doi:10.1140/epjc/s10052-017-4824-7
[arXiv:1611.08182 [hep-th]].

\bibitem{Frasca:2013tma}
M.~Frasca,
Eur. Phys. J. C \textbf{74}, 2929 (2014)
doi:10.1140/epjc/s10052-014-2929-9
[arXiv:1306.6530 [hep-ph]].

\bibitem{Frasca:2012ne}
M.~Frasca,
J. Nonlin. Math. Phys. \textbf{20}, no.4, 464-468 (2013)
doi:10.1080/14029251.2013.868256
[arXiv:1212.1822 [hep-th]].

\bibitem{Frasca:2009bc}
M.~Frasca,
J. Nonlin. Math. Phys. \textbf{18}, no.2, 291-297 (2011)
doi:10.1142/S1402925111001441
[arXiv:0907.4053 [math-ph]].

\bibitem{Frasca:2010ce}
M.~Frasca,
PoS \textbf{FACESQCD}, 039 (2010)
doi:10.22323/1.117.0039
[arXiv:1011.3643 [hep-th]].

\bibitem{Frasca:2008tg}
M.~Frasca,
Nucl. Phys. B Proc. Suppl. \textbf{186}, 260-263 (2009)
doi:10.1016/j.nuclphysbps.2008.12.058
[arXiv:0807.4299 [hep-ph]].

\bibitem{Frasca:2009yp}
M.~Frasca,
Mod. Phys. Lett. A \textbf{24}, 2425-2432 (2009)
doi:10.1142/S021773230903165X
[arXiv:0903.2357 [math-ph]].

\bibitem{Frasca:2008zp}
M.~Frasca,
Int. J. Mod. Phys. E \textbf{18}, 693-703 (2009)
doi:10.1142/S0218301309012781
[arXiv:0803.0319 [hep-th]].

\bibitem{Frasca:2007uz}
M.~Frasca,
Phys. Lett. B \textbf{670}, 73-77 (2008)
doi:10.1016/j.physletb.2008.10.022
[arXiv:0709.2042 [hep-th]].

\bibitem{Frasca:2006yx}
M.~Frasca,
Int. J. Mod. Phys. A \textbf{22}, 2433-2439 (2007)
doi:10.1142/S0217751X07036427
[arXiv:hep-th/0611276 [hep-th]].

\bibitem{Frasca:2005sx}
M.~Frasca,
Phys. Rev. D \textbf{73}, 027701 (2006)
[erratum: Phys. Rev. D \textbf{73}, 049902 (2006)]
doi:10.1103/PhysRevD.73.049902
[arXiv:hep-th/0511068 [hep-th]].

\bibitem{Frasca:2005mv}
M.~Frasca,
Int. J. Mod. Phys. A \textbf{22}, 1441-1450 (2007)
doi:10.1142/S0217751X07036282
[arXiv:hep-th/0509125 [hep-th]].

\bibitem{Frasca:2005fs}
M.~Frasca,
Int. J. Mod. Phys. D \textbf{15}, 1373-1386 (2006)
doi:10.1142/S0218271806009091
[arXiv:hep-th/0508246 [hep-th]].


\bibitem{Frasca:2020jbe}
M.~Frasca and A.~Ghoshal,
Class. Quant. Grav. \textbf{38}, no.17, 17 (2021)
doi:10.1088/1361-6382/ac161b
[arXiv:2011.10586 [hep-th]].

\bibitem{Frasca:2020ojd}
M.~Frasca and A.~Ghoshal,
JHEP \textbf{21}, 226 (2020)
doi:10.1007/JHEP07(2021)226
[arXiv:2102.10665 [hep-th]].

\bibitem{Frasca:2021iip}
M.~Frasca, A.~Ghoshal and N.~Okada,
Phys. Rev. D \textbf{104}, no.9, 096010 (2021)
doi:10.1103/PhysRevD.104.096010
[arXiv:2106.07629 [hep-th]].

\bibitem{Fubini:1976jm}
S.~Fubini,
Nuovo Cim. A \textbf{34}, 521 (1976)
doi:10.1007/BF02785664.

\bibitem{Lipatov:1976ny}
L.~N.~Lipatov,
Sov. Phys. JETP \textbf{45}, 216-223 (1977)
LENINGRAD-76-255.

\bibitem{Pasechnik:2013sga}
R.~Pasechnik, V.~Beylin and G.~Vereshkov,
Phys. Rev. D \textbf{88}, no.2, 023509 (2013)
doi:10.1103/PhysRevD.88.023509
[arXiv:1302.5934 [gr-qc]].

\bibitem{Pasechnik:2013poa}
R.~Pasechnik, V.~Beylin and G.~Vereshkov,
JCAP \textbf{06}, 011 (2013)
doi:10.1088/1475-7516/2013/06/011
[arXiv:1302.6456 [gr-qc]].

\bibitem{Addazi:2018fyo}
A.~Addazi, A.~Marcian\`o, R.~Pasechnik and G.~Prokhorov,
Eur. Phys. J. C \textbf{79}, no.3, 251 (2019)
doi:10.1140/epjc/s10052-019-6780-x
[arXiv:1804.09826 [hep-th]].

\bibitem{Addazi:2018ctp}
A.~Addazi, A.~Marcian\`o and R.~Pasechnik,
Chin. Phys. C \textbf{43}, no.6, 065101 (2019)
doi:10.1088/1674-1137/43/6/065101
[arXiv:1812.07376 [hep-th]].

\bibitem{Batalin:1976uv}
I.~A.~Batalin, S.~G.~Matinyan and G.~K.~Savvidy,
Sov. J. Nucl. Phys. \textbf{26}, 214 (1977)
EFI-198-44-76-YEREVAN.

\bibitem{Savvidy:1977as}
G.~K.~Savvidy,
Phys. Lett. B \textbf{71}, 133-134 (1977)
doi:10.1016/0370-2693(77)90759-6

\bibitem{Pagels:1978dd}
H.~Pagels and E.~Tomboulis,
Nucl. Phys. B \textbf{143}, 485-502 (1978)
doi:10.1016/0550-3213(78)90065-2

\bibitem{Olesen:1980nz}
P.~Olesen,
Phys. Scripta \textbf{23}, 1000-1004 (1981)
doi:10.1088/0031-8949/23/5B/018

\bibitem{Shuryak:1983ni}
E.~V.~Shuryak,
Phys. Rept. \textbf{115}, 151 (1984)
doi:10.1016/0370-1573(84)90037-1

\bibitem{Pasechnik:2016sbh}
R.~Pasechnik,
Universe \textbf{2}, no.1, 4 (2016)
doi:10.3390/universe2010004
[arXiv:1605.07610 [gr-qc]].

\bibitem{Addazi:2022whi}
A.~Addazi, T.~Lundberg, A.~Marcian\`o, R.~Pasechnik and M.~\v{S}umbera,
Universe \textbf{8}, no.9, 451 (2022)
doi:10.3390/universe8090451
[arXiv:2204.02950 [hep-ph]].


\bibitem{Lundow:2022ltb}
P.~H.~Lundow and K.~Markstr\"om,
Nucl. Phys. B \textbf{993}, 116256 (2023)
doi:10.1016/j.nuclphysb.2023.116256
[arXiv:2209.05292 [cond-mat.stat-mech]].


\bibitem{Dunne:2021lie}
G.~V.~Dunne and M.~Meynig,
Phys. Rev. D \textbf{105}, no.2, 2 (2022)
doi:10.1103/PhysRevD.105.025019
[arXiv:2111.15554 [hep-th]].

\bibitem{Bjorken:1965sts}
J.~D.~Bjorken and S.~D.~Drell,
``Relativistic Quantum Mechanics,''
McGraw-Hill, 1965,
ISBN 978-0-07-005493-6, pp.38ff.

\bibitem{Ebert:1982pk}
D.~Ebert and M.~K.~Volkov,
Z. Phys. C \textbf{16}, 205 (1983)
doi:10.1007/BF01571607

\bibitem{Frasca:2013kka}
M.~Frasca,
JHEP \textbf{11}, 099 (2013)
doi:10.1007/JHEP11(2013)099
[arXiv:1309.3966 [hep-ph]].

\bibitem{Frasca:2011bd}
M.~Frasca,
Phys. Rev. C \textbf{84}, 055208 (2011)
doi:10.1103/PhysRevC.84.055208
[arXiv:1105.5274 [hep-ph]].

\bibitem{Rezaeian:2004nf}
A.~H.~Rezaeian, N.~R.~Walet and M.~C.~Birse,
Phys. Rev. C \textbf{70}, 065203 (2004)
doi:10.1103/PhysRevC.70.065203
[arXiv:hep-ph/0408233 [hep-ph]].



\end{thebibliography}
\end{document}